\documentclass[preprintnumbers,article,amsmath,amssymb,floatfix,10pt,prd,twocolumn,
superscriptaddress,nofootinbib]{revtex4-2}
\usepackage{bm}
\usepackage{amsfonts}
\usepackage{latexsym}
\usepackage[latin1]{inputenc}
\usepackage{graphicx}
\usepackage{amsmath}
\usepackage{palatino}
\usepackage{mathpazo}
\usepackage{textcomp}
\usepackage{float}
\usepackage{booktabs}
\usepackage{dcolumn}
\usepackage{ragged2e}
\usepackage{hyperref}
\hypersetup{colorlinks,citecolor=blue}
\hypersetup{colorlinks=true,linkcolor=red,filecolor=magenta, urlcolor=blue}
\usepackage{amsmath}
\usepackage{xcolor}
\usepackage{orcidlink}
\usepackage{epsfig}
\usepackage[caption=false]{subfig}
\usepackage{commath}
\usepackage{tabularx}
\usepackage{multirow}
\usepackage{mathtools}
\newcommand{\be}{\begin{equation}}
\newcommand{\ee}{\end{equation}}
\newcommand{\bea}{\begin{eqnarray}}
\newcommand{\eea}{\end{eqnarray}}
\newcommand{\eeas}{\end{eqnarray*}}
\newcommand{\beas}{\begin{eqnarray*}}

\def\jnl@style{\it}
\def\aaref@jnl#1{{\jnl@style#1}}

\def\aaref@jnl#1{{\jnl@style#1}}

\def\aj{\aaref@jnl{AJ}}                   
\def\apj{\aaref@jnl{ApJ}}                 
\def\apjl{\aaref@jnl{ApJ}}                
\def\apjs{\aaref@jnl{ApJS}}               
\def\apss{\aaref@jnl{Ap\&SS}}             
\def\aap{\aaref@jnl{A\&A}}                
\def\aapr{\aaref@jnl{A\&A~Rev.}}          
\def\aaps{\aaref@jnl{A\&AS}}              
\def\mnras{\aaref@jnl{Mon.~Not.~Roy.~Astron.~Soc.}}             
\def\prd{\aaref@jnl{Phys.~Rev.~D}}        
\def\prc{\aaref@jnl{Phys.~Rev.~C}}  
\def\prl{\aaref@jnl{Phys.~Rev.~Lett.}}    
\def\qjras{\aaref@jnl{QJRAS}}             
\def\skytel{\aaref@jnl{S\&T}}             
\def\ssr{\aaref@jnl{Space~Sci.~Rev.}}     
\def\zap{\aaref@jnl{ZAp}}                 
\def\nat{\aaref@jnl{Nature}}              
\def\aplett{\aaref@jnl{Astrophys.~Lett.}} 
\def\apspr{\aaref@jnl{Astrophys.~Space~Phys.~Res.}} 
\def\physrep{\aaref@jnl{Phys.~Rep.}}      
\def\physscr{\aaref@jnl{Phys.~Scr}}       
\def\commat{\aaref@jnl{Comm.~Math.~Phys.}}              
\def\science{\aaref@jnl{Science}}               
\def\cqg{\aaref@jnl{Classical Quant.~Grav.}}            
\def\jpcs{\aaref@jnl{JPCS}}                                     
\def\ijmpd{\aaref@jnl{Int.~J.~Mod.~Phys.~D}}                    
\def\grg{\aaref@jnl{Gen.~Relat.~Gravit.}}               
\def\rpp{\aaref@jnl{Rep.~Prog.~Phys.}}          
\def\npa{\aaref@jnl{Nucl.~Phys.~A}}        
\def\lrr{\aaref@jnl{Living Rev.~Rel.}}                   
\def\jcap{\aaref@jnl{J.~Cosmology Astropart.~Phys.}}    
\def\rmp{\aaref@jnl{Rev.~Mod.~Phys.}}   
\def\epjc{\aaref@jnl{Eur.~Phys.~J.~C}} 
\def\plb{\aaref@jnl{~Phy.~Lett.~B}} 
\def\mpla{\aaref@jnl{Mod.~Phy.~Lett.~A}} 
\def\arxiv{\aaref@jnl{arxiv.org}}

\allowdisplaybreaks[1]

\begin{document}
\color{black}       
\title{Barrow Holographic Dark Energy in $f(Q,T)$ gravity}

\author{ N. Myrzakulov\orcidlink{0000-0003}}\email{nmyrzakulov@gmail.com}
\affiliation{L N Gumilyov Eurasian National University, Astana 010008, Kazakhstan}
\author{ S. H. Shekh\orcidlink{0000-0002-1932-8431}}\email{  da\_salim@rediff.com}
\affiliation{L N Gumilyov Eurasian National University, Astana 010008, Kazakhstan}
\affiliation{Department of Mathematics, S.P.M. Science and Gilani Arts, Commerce College, Ghatanji, Yavatmal, \\Maharashtra-445301, India.}

\author{Anirudh Pradhan\orcidlink{0000-0002-1932-8431}}
\email{pradhan.anirudh@gmail.com}
\affiliation{Centre for Cosmology, Astrophysics and Space Science (CCASS), GLA University, Mathura-281406, U.P., India.}

\author{K Ghaderi
}
\email{k.ghaderi.60@gmail.com}
\affiliation{Department of Physics, Marivan Branch, Islamic Azad University, Marivan, Iran} 
\begin{abstract}
\textbf{Abstract:} 
In the present analysis, we explore a new version of dark energy called Barrow holographic dark energy within the framework modified gravity called $f(Q,T)$ gravity by adopting the simple homogeneous, isotropic, and spatially flat Friedmann-Robertson-Walker (FRW) model of the universe. Our goal is to understand how the universe evolved over time. To do this, we use parameterizetion of Hubble's parameter method. We then use a powerful tool called Monte Carlo Markov Chain to find the best values for the constants in our formula. We do this by comparing our formula to actual data from observations of the universe. Once we have the best values for the constants, we calculate other important parameters that describe the universe's evolution. These include: Deceleration parameter which measures how quickly the expansion is slowing down. We found $q_0 = -0.601^{+0.0131}_{-0.0131}$. Equation of state parameter to measures the properties of dark energy. We find $\omega_0 = -0.7018^{+0.0101}_{-0.0101}$.
We also study the stability and energy conditions along with the state-finder and $O_m(z)$-parameter of our model to ensure it's consistent with our understanding of the universe.


\end{abstract}

\maketitle

\ 

\section{introduction}
Einstein's standard $\Lambda$CDM model has been remarkably consistent with the majority of astrophysical and cosmological observational data conducted over the past decades \cite{1,2,3,4,5,6,7}. Nevertheless, when analyzing different datasets found certain discrepancies from the standard model, such as the Hubble tension \cite{8,9} reaching to $5\sigma$ level of significance \cite{10,11,12} and the $S_{8}$ tension reaching $3\sigma$ \cite{13,14,15,16,17,18,19,20} in the new era of high-precision cosmology. This pivotal situation has compelled the scientific community to embark on a quest for the identification of potential/alternative explanations rooted in novel physics. 

Many existing issues in studying the dynamics and nature of our Universe have led to two main approaches to its description. The first step maintains general relativity and introduces the concept of dark energy, which can account for all forms of new/exotic sectors that can be sources of acceleration phenomenon \cite{21,22,23,24,25}. The second approach is to attribute the new degrees of freedom to modifications of the gravitational interaction, namely to modified theories  \cite{26,27}. The idea to modify General Relativity theory on cosmological scales has taken off over the past decade. This has been interesting by the scientific community, in part, by theoretical developments involving higher dimensional theories, as well as new developments in constructing renormalizable gravity theories. 

Many recent theoretical investigations have indicated the important result that general relativity can be described in three mathematical geometric formalisms that are equivalently between them, namely curvature, torsion, and non-metricity. Over the past decade, modified $f(R)$ gravity in terms of curvature have been extensively studied as one of the simplest modifications to General Relativity and has attracted the interest of many cosmologists \cite{28}. The discovery of the cosmic acceleration phenomenon stimulated the idea of studying different dark energy models based on $f(R)$ theories of gravity. Another alternative modification in terms of torsion (teleparallel formulation) with curvature vanishing is called the $f(T)$ gravity theory \cite{29,30}. The $f(T)$ gravity has interesting cosmological solutions, which provide alternative interpretations for the accelerating expansion phases of the Universe. Modified gravity known as $f(Q)$ gravity in terms of non-metricity has gained considerable popularity in the past years and research efforts have been dedicated to cosmological applications \cite{31,32}. $f(Q)$ gravity extensively has been applied to reconstruct of $\Lambda$CDM universe \cite{33}, quantum cosmology \cite{34}, to find wormhole solutions \cite{35} as well as static and spherically symmetric solutions \cite{36}, to study late-time cosmology with phantom dark energy \cite{37}, bouncing cosmological scenario \cite{38}, slow-roll inflation \cite{39}. 

Our research focus is to study novel extension of the $f(Q)$ gravity as $f(Q,T)$ where the Lagrangian $L$ is represented through an arbitrary function of the non-metricity $Q$ and the trace of the energy-momentum tensor $T$ \cite{40}.  This kind of modified gravity can provide a promising framework to evamine the dynamics of the Universe on cosmological scales. Many works have been presented on $f(Q,T)$ gravity in the literature that can provide useful insights for the description of the early and late phases of cosmological evolution. Specifically, Ref. \cite{40} presents the modified Einstein-Hilbert action for extended symmetric teleparallel gravity, which incorporates this $f(Q, T)$ coupling as,
\begin{equation}\label{e1}
	S=\int \left[ \frac{1}{16\pi }f(Q,T)+L_{m}\right] \sqrt{-g}d^{4}x,
\end{equation}%
where $g$ is the metric tensor's determinant $g_{\mu \nu }$ i.e. $g=\det \left( g_{\mu \nu }\right) $, $f(Q,T)$ is the generic function of the non-metricity scalar $Q$ and its trace, $T$ and $L_{m}$ is the usual matter Lagrangian.\\ 
The non-metricity scalar $Q$ is
defined as
\begin{equation}\label{e2}
	Q\equiv -g^{\mu \nu }\left( L^{\delta }{}_{\alpha \mu }L^{\alpha }{}_{\nu
		\delta }-L^{\delta }{}_{\alpha \delta }L^{\alpha }{}_{\mu \nu }\right) ,
\end{equation}%
where $L^{\delta }{}_{\alpha \gamma }=-\frac{1}{2}g^{\delta \eta }\left( \nabla
_{\gamma }g_{\alpha \eta }+\nabla _{\alpha }g_{\eta \gamma }-\nabla _{\eta
}g_{\alpha \gamma }\right)$, $Q_{\gamma \mu \nu }=\nabla _{\gamma }g_{\mu \nu }$ and 
$Q_{\delta }=g^{\mu \nu }Q_{\delta \mu \nu },\text{ \ \ \ \ }\widetilde{Q}%
_{\delta }=g^{\mu \nu }Q_{\mu \delta \nu }$ are respective the dis-formation tensor, non-metricity tensor, and the trace of the non-metricity tensor.\\
In addition, the field equations of $f\left( Q,T\right) $ gravity are obtained by modifying the action, Eq. (\ref{e1}), in relation to the metric tensor $g_{\mu \nu }$
\begin{widetext}
	\begin{equation}\label{e6}
		-\frac{2}{\sqrt{-g}}\nabla _{\delta }\left( f_{Q}\sqrt{-g}P^{\delta }{}_{\mu
			\nu }\right) -\frac{1}{2}fg_{\mu \nu }+f_{T}\left( T_{\mu \nu }+\theta _{\mu
			\nu }\right) -f_{Q}\left( P_{\mu \delta \alpha }Q_{\nu }{}^{\delta \alpha
		}-2Q^{\delta \alpha }{}_{\mu }P_{\delta \alpha \nu }\right) =8\pi T_{\mu \nu
		},
	\end{equation}
\end{widetext}
where $f_{Q}=\frac{df\left( Q,T\right) }{dQ}$, $f_{T}=\frac{df\left(
	Q,T\right) }{dT}$, and the covariant derivative is denoted by the symbol $\nabla _{\delta }$. According to Eq. (\ref{e6}), the tensor $\theta _{\mu \nu }$ affects how the field equations of $f(Q,T)$ extended symmetric teleparallel gravity function. \\

 In \cite{41} developed the cosmological linear (tensor, vector, scalar) perturbations for $f(Q,T)$ gravity and study the effects of the coupling between the trace of the stress energy tensor $T$ and the non-metricity scalar $Q$ by considering the weak and strong coupling limits. Cosmology of holographic origin within modified $f(Q,T)$ gravity framework by employing different forms of the scale factor \cite{42}. Fundamental energy conditions are investigated by using the proposed deceleration parameter, which predicts both decelerated and accelerated phases of the Universe in context of $f(Q,T)$ gravity \cite{43}. Gravastar with three different zones have been examined and presented a stable gravastar model in \cite{44}. Behavior of different cosmological parameters and comparison them with the observational data explored in \cite{45}.

In addition to the above, several other theoretical ways to the DE problem have been developed. In addition, the important way to apply a holographic principle is that the universe horizon entropy is proportional to its area (similar to the Bekenstein-Hawking entropy of a black hole). Holographic dark energy is an alternative quantitative description of dark energy, originating from the holographic principle \cite{46,47}. Recently Barrow constructed new holographic dark energy by using the extended relation for the horizon entropy, instead of the usual Bekenstein-Hawking one \cite{48}. In \cite{49} formulated Barrow holographic dark energy, by applying the usual holographic principle at a framework of cosmology. Generalized interacting model of Barrow Holographic Dark Energy with infrared cutoff constructed in \cite{50}. In \cite{51} studied the warm inflation mechanism with the Barrow holographic dark energy model. Barrow holographic dark energy are studied in the context of Randall and Sundrum (RS II) brane-world, Dvali-Gabadadze-Porrati (DGP) brane, and the cyclic Universe in \cite{52}. Investigation on the effects of implementing the Granda-Oliveros infrared cutoff in Barrow holographic dark energy framework \cite{53}. Recently, a number of authors \cite{54,55,56,57,58,59} have conducted research on the Barrow holographic dark energy models in a variety of contexts.

Manuscript organized as following structure: Section II: We delved into the field equation and Barrow holographic dark energy, laying the groundwork for our analysis. Section III: We focused on a specific form of the Hubble parameter ($H(z)$) and examined some cosmographic parameters, which help us understand the universe's evolution. Section IV: We analyzed the physical aspects of the model, discussing its implications on the universe's expansion and properties. Section V: We concluded our findings, summarizing the key points and takeaways from our analysis.

\section{Field equations and Barrow HDE}

To facilitate the solution of field equations in $f(Q,T)$ extended symmetric teleparallel gravity, simplifying assumptions are often necessary. In this work, we adopt the homogeneous, isotropic, and spatially flat Friedmann-Robertson-Walker (FRW) metric, given by:
\begin{equation}\label{e8}
	ds^{2}=-dt^{2}+\delta_{ij} g_{ij} dx^{i} dx^{j},{\;\;\;\;} i,j=1,2,3,.....N,
\end{equation}
This choice of metric allows us to explore the cosmological implications of $f(Q,T)$ gravity in a straightforward and analytically tractable manner.
where $g_{ij}$ are the function of $(-t, x^{1}, x^{2}, x^{3})$ and $t$ refers to the cosmological/cosmic time measure in Gyr. In four dimensional FRW space-time, from above equation we have
\begin{equation}\label{e9}
	\delta_{ij} g_{ij}=a^{2}(t,x)
\end{equation} 
where $a$ be the average scale factor of the Universe and $t$ is the cosmic time in Gyr. Above relations show that for FRW universe all three metric are equal (i.e $g_{11} = g_{22} = g_{33} =a^{2}(t, x)$). 
Thus, the corresponding non-metricity scalar for the line element  Eq. (\ref{e8}) is given by $Q=6H^{2}$, where $H$ be the average Hubble's parameter of the form $H=\frac{\dot{a}}{a}$.\\
We consider a perfect fluid for which: 
\begin{equation}\label{e13}
T_{\nu }^{\mu }=diag\left( -\rho ,p,p,p\right) , 
\end{equation}%
where $\rho $ is the energy density and the isotropic pressure is $p$. The field equations (\ref{e6}) for the metric
(\ref{e8}) yield
\begin{equation}\label{e14}
\kappa^{2} \rho =\frac{f}{2}-6FH^{2}-\frac{2\widetilde{G}}{1+\widetilde{G}}\left( 
\overset{.}{F}H+F\overset{.}{H}\right) ,  
\end{equation}
\begin{equation}\label{e15}
\kappa^{2} p=-\frac{f}{2}+6FH^{2}+2\left( \overset{.}{F}H+F\overset{.}{H}\right) .
\end{equation}
where, where $\kappa^{2} \widetilde{G}\equiv f_{T}$ (here $\kappa^{2}=1$) and $F\equiv f_{Q}$  are derivatives with respect to  $T$ and $Q$,  respectively. The Hubble parameter $H$ is given by $H\equiv \dot{a}/{a}$ and $\left( \text{\textperiodcentered }\right) $ is $d/dt$. \\

Barrow proposed a new idea that the entropy of a black hole (a measure of its internal disorder) follows a more universal relationship: 
\begin{equation}
	S_{h}=\left( \frac{A}{A_{0}}\right) ^{(1+\bigtriangleup /2)},  \label{e1}
\end{equation}%
where $A$ and $A_{0}$ are the area of the black hole horizon and plank area respectively. $\bigtriangleup $ be the exponent which computes the extent of quantum-gravitational deformation effects. The standard holographic dark energy model is based on the idea that the energy density of the universe is related to its horizon area. Barrow modified this idea to create a new model, called Barrow holographic dark energy, which is described by the equation: 
\begin{equation}
	\rho _{B}=CL^{\Delta -2},  \label{e2}
\end{equation}%
where $C$ is a parameter with dimension $\left[ L\right] ^{-2-\Delta }$, $L$
can be considered as the size of the current Universe such as the Hubble's
scale and the future event horizon, and $\Delta $ is a free parameter.

Numerous physical parameters or attributes within a cosmology are closely associated with the above energy density and isotropic pressure, and their behavior may frequently be studied by analyzing their expressions or interpreting their graphical representations. The following will be the examination of the expressions of certain important components, such as the equation of state parameter, 
and the energy conditions.\\
\textit{The Equation of State parameter is}
\begin{equation}\label{e16}
\omega=\frac{p}{\rho}
\end{equation}
Interestingly, the final nature of DE is often classified using the so-called Equation of State (EoS) parameter, which quantifies the correlation between spatially homogeneous pressure and energy density. We may now comprehend the importance of the EoS parameter, $\omega < -1/3$, which is necessary for rapid cosmic expansion according to recent cosmological investigations. The most significant options in this classification are scalar field models with an EoS value of $-1<\omega <-1/3$, also referred to as a Quintessence field DE 
, as opposed to $\omega <-1$, which is called a phantom field DE 
. Additionally, the EoS parameter for DE is now $\omega_0 = -1.084 \pm 0.063$ according to the combined observations of WMAP9 and the $H_0$ measurements, Supernovae of type Ia (SNe Ia), CMB, and BAO (Baryon Acoustic Oscillations). Also, we improved $\omega_0 = -1.028^{+0.032}_{-0.032}$ in 2018 and the Planck collaboration found in 2015 that $\omega_0 = -1.006^{+0.0451}_{-0.0451}$.
 \\

\section{ Specific $H(z)$ and some Cosmographic  parameters}
Within the framework of symmetric teleparallel gravity, we have assumed a perfect fluid as the matter content of the Universe. To determine the expansion rate, we define the dimensionless function $E(z)$ as \cite{60}:
$$E(z) = \frac{H^2(z)}{H_0^2} = \Omega_{0m}(1+z)^3 + \Omega_{0_{Q,T}}$$
where $H(z)$ is the Hubble parameter at redshift $z$, $H_0$ is the present-day value of the Hubble constant, and $\Omega_{0m}$ is the present-day value of the matter energy density parameter. Note that $\Omega_{0_{Q,T}}$ represents the energy density parameter arising from the geometry of $f(Q,T)$ gravity. An alternative functional form of $E(z)$ can be expressed as Mahmood et al. \cite{54} $$	E(z) = \Omega_{0m}(1+z)^{3} + \alpha(1+z)^{2} + \beta(1+z) + \mu   $$
It is important to note that at $z = 0$, the Hubble parameter $H(z)$ equals the present-day value $H_0$, which implies that $E(z) = 1$ at $z = 0$. This constraint validates above Eq. $\alpha$, $\beta$, and $\mu$ to $\alpha + \beta + \mu = 1 - \Omega_{0m}$. To satisfy this condition, we consider the simplest functional form as $\Omega_{0_{Q,T}} = 1 - \Omega_{0m}$. Consequently, we can rewrite the above equation as \cite{54} :
\begin{equation}
	H(z) = H_0\left[\Omega_{0m}(1+z)^3 + (1-\Omega_{0m})\right]^{\frac{1}{2}}
\end{equation}
This expression represents the Hubble parameter $H(z)$ in terms of the redshift $z$, the present-day matter energy density parameter $\Omega_{0m}$, and the present-day Hubble constant $H_0$.\\
We introduced a new way to measure the deceleration parameter, which helps us understand the universe's expansion. To find the best fit, we used two types of data:

1. Cosmic Chronometer (OHD) data: 55 data points measuring the universe's expansion history up to a certain point ($0 \le z \le 2.36$) whose $\chi^2_{OHD}$ is defined as 
\begin{equation}\label{e19}
	\chi_{OHD}^2=\sum_{i=1}^{55} \frac{\left[H_{t h}-H_{o b
			s}\left(z_i\right)\right]^2}{\sigma_{H\left(z_i\right)}^2},
\end{equation}
2. Standard Candles (SN Ia) data from Pantheon: 1048 data points measuring the brightness of supernovae explosions up to a certain point ($0.01 \le z \le 2.36$) whose $\chi^2_{SN}$ is defined as 

We combined these data sets to constrain our model using a statistical method called MCMC. We minimized the total $\chi^2$ function to validate our model with the available data. The total $\chi^2$ function is defined as
\begin{equation}\label{e118}
	\chi^2=\chi_{OHD}^2+\chi_{ SN }^2
\end{equation}

\begin{figure}[H]
	\centering
	\includegraphics[scale=0.6]{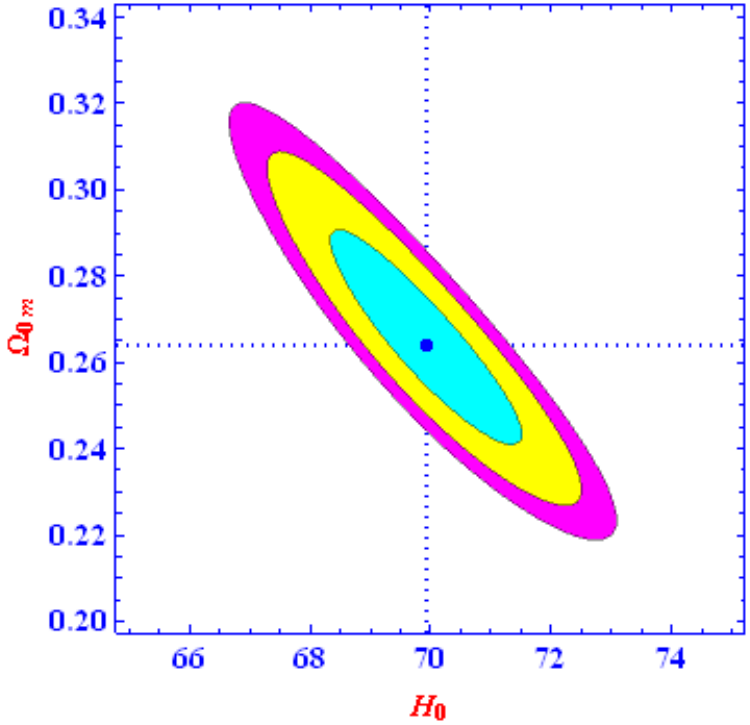}
	\caption{Above figure shows the combined visualization of two dimensional contours at 1$\sigma$ \& 2$\sigma$ confidence regions by bounding our model with OHD data sets and OHD + Pantheon compilation of SN Ia data}\label{HP}
\end{figure}

Figure \ref{HP}   shows the two-dimensional contour plots based on the combined data sets. The contours represent the confidence levels of our model's parameters, showing how well they fit the data.
\subsubsection{Deceleration parameter}
As is well established, the universe's initial expansion was decelerating due to the strong gravitational attraction between matter and radiation. However, as the universe expanded and matter became increasingly dispersed, the gravitational force weakened, leading to a shift in cosmic dynamics. The universe has now transitioned to a phase of accelerated expansion, characterized by a negative deceleration parameter ($q$). Investigating this transition is essential for understanding the underlying mechanisms driving the universe's expansion. The deceleration parameter ($q$) provides a quantitative measure of this acceleration, enabling us to evaluate the current state of cosmic expansion. Specifically, the deceleration parameter ($q$) is defined in terms of the Hubble parameter ($H$) as:
\begin{equation}
	q(z) = -1+\frac{(1+z)}{H(z)}\frac{dH(z)}{dz}.
\end{equation}

\begin{figure}[H]
		\centering
		\includegraphics[scale=0.7]{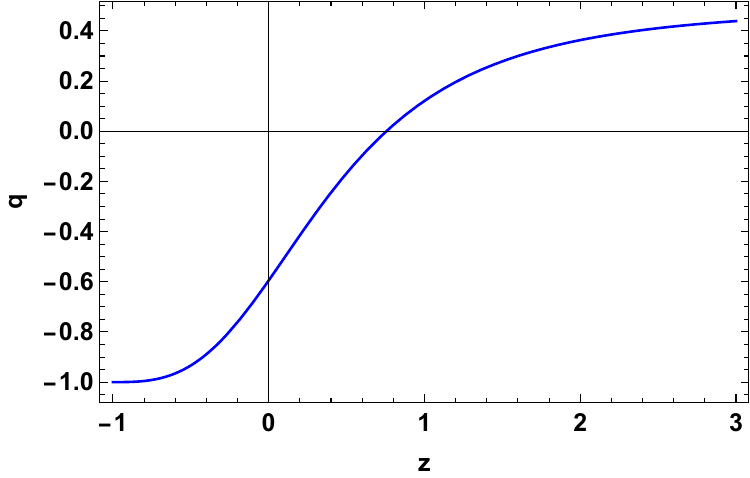}
		\caption{Above figure shows the behavior of $q(z)$ with the  constraint values of the cosmological free parameters}\label{q}
\end{figure}

The sign of the deceleration parameter ($q$) serves as a diagnostic tool for distinguishing between decelerating and accelerating phases of cosmic expansion, with positive values indicating deceleration and negative values signifying acceleration. Recent cosmological observations have precisely constrained the present-day value of $q$ to $q_0 = -0.51^{+0.09}_{-0.01}$ \cite{61} and $q_0 = -0.5422_{-0.0826}^{+0.0718}$ \cite{62}, providing robust evidence for the current accelerating expansion of the Universe. Moreover, the transition redshift ($z_t$) from deceleration to acceleration has been measured to be $z_t = 0.65^{+0.19}_{-0.17}$ and $z_t = 0.8596^{+2886}_{-0.2722}$, marking a critical epoch in the Universe's evolution \cite{63}.For the considered Hubble parameter, the deceleration parameter $q(z)$ is observed to exhibit the following behavior:
\begin{equation}
	q(z) = -1+\frac{3 \Omega_{0m} (z+1)^3}{2 \Omega_{0m} z (z (z+3)+3)+2}.
\end{equation}
The aforementioned phase transition and evolutionary behavior of the deceleration parameter are vividly illustrated in Fig. \ref{q}, providing a clear visual representation of the transition from deceleration to acceleration at $z = 0.7413^{+0.0012}_{-0.0012}$ and the present-day value of $q_0 = -0.601^{+0.0131}_{+0.0131}$.
Notably, our investigation reveals a phase transition of the deceleration parameter from deceleration to acceleration at a redshift value of $z = 0.7413^{+0.0012}_{-0.0012}$. Moreover, the present-day value of the deceleration parameter is determined to be $q_0 = -0.6$, which is in excellent agreement with the cosmological observations reported in Ref. \cite{61,62,63}. This consistency reinforces the validity of our findings and provides further evidence for the accelerating expansion of the Universe.\\

\subsubsection{State-finder parameters}
The pivotal role of dark energy in fueling the accelerating cosmic expansion is well-established, prompting intense research interest in recent decades. This has led to the development of a plethora of dark energy models, each attempting to elucidate the enigmatic nature of dark energy. To differentiate between these models, both quantitatively and qualitatively, Sahni et al. \cite{64} introduced a pair of geometrical parameters, known as statefinder parameters ($r$, $s$), which are defined as:

 \begin{eqnarray}
	j(z) &=& q(z) + 2q^{2}(z) + (1+z)\frac{dq(z)}{dz},\label{eq:27} \\
	s(z) &=& \frac{j(z)-1}{3\left(q(z)-\frac{1}{2}\right)}, ~~~~~~~~~~~~~~\left( q\neq \frac{1}{2}\right).
	\label{eq:28}
\end{eqnarray}
where $a$ is the scale factor, $H$ is the Hubble parameter, and $q$ is the deceleration parameter. These statefinder parameters offer a robust diagnostic tool for distinguishing between various dark energy models, enabling researchers to constrain and characterize the properties of dark energy. The authors present in Ref. \cite{64,65,66} have categorized various dark energy models based on the state-finder parameters such as $r=1,s=0  \rightarrow \Lambda$CDM; $r<1, s>0 \rightarrow$  Quintessence;
$j>1,s<0 \rightarrow $ Chaplygin Gas;
$j=1,s=1 \rightarrow$ SCDM.
Notably, the point $\{r,s\} = \{1, 0\}$ serves as a reference point, representing the flat $\Lambda$CDM model. For the considered Hubble parameter, the deceleration parameter $q(z)$ is observed as:

\begin{equation}
	r=1 {\;\;\;}\text{and}{\;\;\;\;} s=0
\end{equation}
The statefinder parameters $r$ and $s$ exhibit constant values, indicating that the cosmological model under consideration is characterized by a fixed evolutionary trajectory. The statefinder parameters $r=1$ and $s=0$ correspond to the $\Lambda$CDM (Lambda-Cold Dark Matter) model, which is a standard model of cosmology that describes the evolution of the universe on large scales. In this model, the dark energy is represented by a cosmological constant ($\Lambda$) that drives the accelerating expansion of the universe.

The value of $r=1$ indicates that the expansion history of the universe is consistent with the $\Lambda$CDM model, while $s=0$ suggests that the equation of state of dark energy is consistent with a cosmological constant, i.e., $\omega = -1$.

\subsubsection{$O_m(z)$ Diagnostic:}
The $O_m(z)$ diagnostic provides a novel approach to cosmological analysis, offering a robust alternative to traditional methods. This diagnostic tool has been shown to effectively distinguish between a wide range of dark energy models, including quintessence and phantom models, due to its high sensitivity to the equation of state (EoS) parameter. Extensive studies in the literature have demonstrated the efficacy of $O_m(z)$ in constraining dark energy models  $O_m(z)$ diagnostic is defined as:
\begin{equation}
	O_m(z) = \frac{\left[\frac{H(z)}{H_{0}}\right]^2 - 1}{(1+z)^{3}-1}, \hspace{0.5cm}
\end{equation}
where $H(z)$ is the Hubble parameter at redshift $z$, and $H_0$ is the present-day Hubble constant. This diagnostic has emerged as a powerful tool for probing the properties of dark energy and understanding the evolution of the universe. For the considered Hubble parameter, the  $O_m(z)$ diagnostic parameter is observed as:
\begin{equation}
	O_m(z) =\Omega_{0m}=0.270
\end{equation}

In our investigation, the $O_m(z)$ diagnostic parameter exhibits a constant value, indicating that the cosmological model under consideration is characterized by a flatness in the $O_m(z)$ curve. This constancy suggests that the model is consistent with the $\Lambda$CDM paradigm, where the dark energy density is proportional to the critical density. A constant $O_m(z)$ value also implies that the universe's expansion history is well-described by a single component, such as a cosmological constant, with no evidence for deviations or evolution in the dark energy's properties.

\section{Physical aspects and analysis of the model}

In this section, We commence our investigation by examining the inaugural $f(Q,T)$ gravity model, a groundbreaking framework that extends general relativity by incorporating the non-metricity scalar $Q$ and the trace of the energy-momentum tensor $T$. This foundational model proposes a linear functional form, $f(Q,T)=\alpha Q+ \beta T$, where $\alpha$ and $\beta$ are dimensionless constants that govern the strength of non-metricity and matter-energy interactions, respectively. The term $\alpha Q$ captures the effects of non-metricity, quantifying deviations from Riemannian geometry, while $\beta T$ represents the contribution from the matter-energy sector, encompassing cosmic fluid dynamics. By scrutinizing this elementary model, we can uncover the fundamental implications of non-metricity and its interplay with matter-energy density, providing a foundation for understanding the universe's more complex dynamics. For the said model of $f(Q,T)$ gravity and the Hubble's parameter, from equations (\ref{e15}) - (\ref{e2} ) We derive the cosmographic parameters such as Energy density, pressure, Equation of state parameter, Stability parameter, Energy conditions and discuss their implications, revealing insights into the universe's dynamics and evolution.
\subsubsection{Energy density}
\begin{figure}[H]
	\centering
	\includegraphics[scale=0.7]{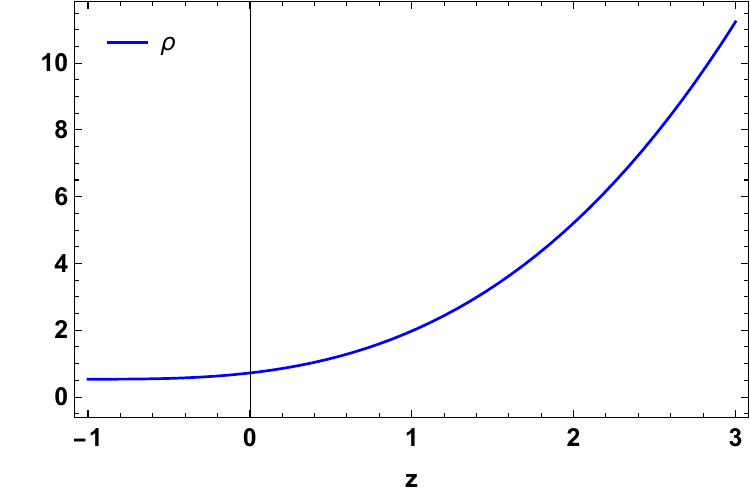}
	\caption{Above figure shows the behavior of $q(z)$ with the  constraint values of the cosmological free parameters}\label{den}
\end{figure}

The energy density of the model is obtained as
\begin{equation}
	\rho = C \left(H_0 \sqrt{\Omega_{0m}  (z (z+3)+3)z+1}\right)^{2-\Delta}
\end{equation}
The cosmic energy density, as prescribed by the $f(Q,T)$ gravity model, is mathematically encapsulated in equation (36). Its evolutionary trajectory with respect to redshift $z$ is visually depicted in Fig. \ref{den},  $\rho$ with increasing $z$. As $z$ approaches asymptotic limits ($z \rightarrow \infty$), the energy density diverges to infinity ($\rho \rightarrow \infty$), implying a singularity-like state in the universe's primordial epoch. Conversely, as $z$ approaches the cosmological horizon ($z \rightarrow -1$), the energy density asymptotes to a finite, positive value, indicative of an expanding universe. These findings offer profound insights into the cosmogonic evolution of the universe within the theoretical framework of this $f(Q,T)$ gravity model.
\subsubsection{Pressure}
The pressure of the model is obtained as
\begin{equation}
	p=-\frac{3 \alpha  H_0^2 \left(2 (\Omega_{0m}-1)+\beta  \left(\Omega_{0m} \left(z^3+3 z^2+3 z+3\right)-2\right)\right)}{4 \beta ^2+6 \beta +2}
\end{equation}

\begin{figure}[H]
	\centering
	\includegraphics[scale=0.7]{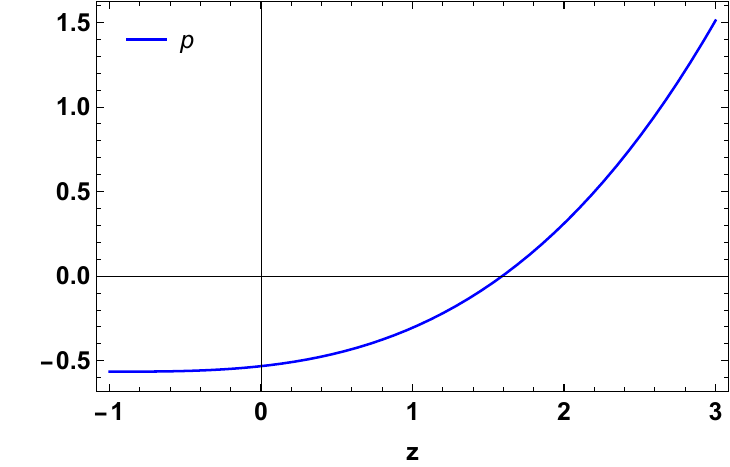}
	\caption{Above figure shows the behavior of $q(z)$ with the  constraint values of the cosmological free parameters}\label{p}
\end{figure}

Cosmological observations imply that dark energy drives the universe's accelerating expansion, with negative pressure serving as a hallmark of its existence. The pressure evolution in our model, $p(z)$, is illustrated in Fig. \ref{p}, $z$ approaches infinity, pressure asymptotes to infinity ($p \rightarrow \infty$ as $z \rightarrow \infty$), indicating a singular behavior in the distant past. Remarkably, pressure transitions to negative values in the current epoch ($z=0$) and persists in the future ($z<0$), aligning with the anticipated characteristics of dark energy. This behavior lends additional credence to our $f(Q,T)$ gravity model as a viable descriptor of the universe's accelerating expansion.
\subsubsection{Equation of state parameter}
\begin{figure}[H]
	\centering
	\includegraphics[scale=0.7]{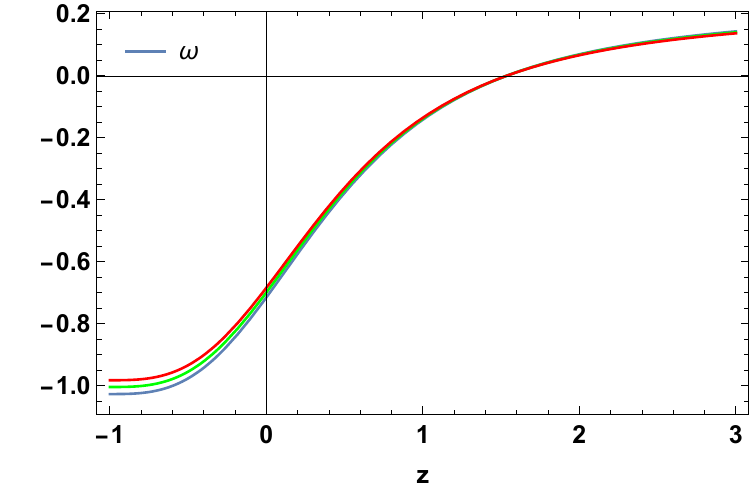}
	\caption{Above figure shows the behavior of $q(z)$ with the  constraint values of the cosmological free parameters}\label{w}
\end{figure}

The pressure of the model is obtained as
\begin{widetext}
\begin{equation}
	\omega = -\frac{3 \alpha  (\Omega_{0m} (\beta  (z (z (z+3)+3)+3)+2)-2 (\beta +1)) \left(H_0 \sqrt{\Omega_{0m} z (z (z+3)+3)+1}\right)^\Delta}{2 C (\beta +1) (2 \beta +1) (\Omega_{0m} z (z (z+3)+3)+1)}
\end{equation}
\end{widetext}
The cosmological trajectory of the equation of state (EoS) parameter $\omega$ as a function of redshift $z$ is depicted in Fig. \ref{w}, ($\omega>0$) at high redshifts, followed by a transition to a quintessence-dominated era ($\omega>-1$) in the present epoch. The EoS parameter asymptotically converges towards the cosmological constant boundary ($\omega=-1$) in the distant future. This evolutionary sequence signifies a paradigm shift from a decelerating to an accelerating expansion phase of the universe. Remarkably, the contemporary value of the EoS parameter, $\omega_0 = -0.7018^{+0.0101}_{-0.0101}$, unequivocally indicates that the universe is presently experiencing an accelerated expansion. These results offer profound insights into the dynamical evolution of the cosmos within the theoretical framework of our $f(Q,T)$ gravity model.
\subsubsection{Stability parameter}
\begin{figure}[H]
	\centering
	\includegraphics[scale=0.7]{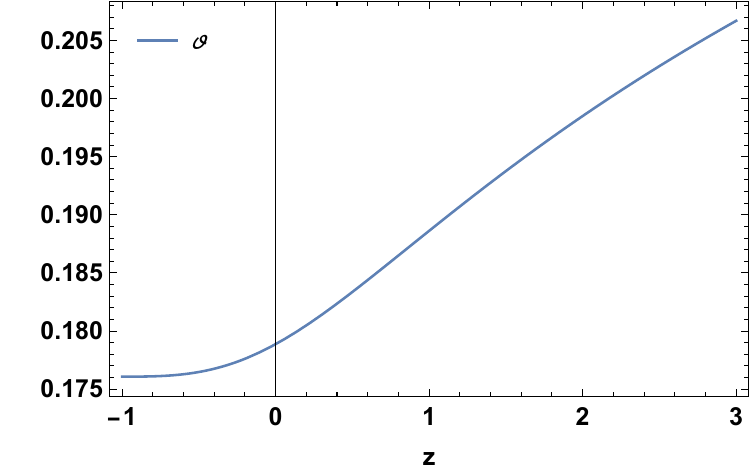}
	\caption{Above figure shows the behavior of $q(z)$ with the  constraint values of the cosmological free parameters}\label{v}
\end{figure}

We examined the stability of the Barrow holographic dark energy model within the $f(Q,T)$ gravity framework by analyzing the squared speed of sound. Hence, the squared speed of sound parameter of the model is obtained as
\begin{equation}
\vartheta^2=	\frac{3 \alpha  \beta  \left(H_0 \sqrt{\Omega_{0m}  (z (z+3)+3)z+1}\right)^\Delta}{C(\beta +1) (2 \beta +1)  (\Delta-2)}
\end{equation}
Our results indicate that $\vartheta^2$ remains positive throughout the universe's evolution, signifying stability of the model against perturbations also exhibits a decreasing trend over time, implying a gradual reduction in the model's stability as the universe expands (See Fig. \ref{v}). However, the values of $\vartheta^2$ remain within reasonable ranges, ensuring the model's overall stability and viability in describing the universe's evolution. Our stability analysis reinforces the credibility of the Barrow holographic dark energy model within the $f(Q,T)$ gravity framework, providing further evidence for its potential to accurately describe the cosmos.

\subsubsection{Energy conditions}
\begin{figure}[H]
	\centering
	\includegraphics[scale=0.7]{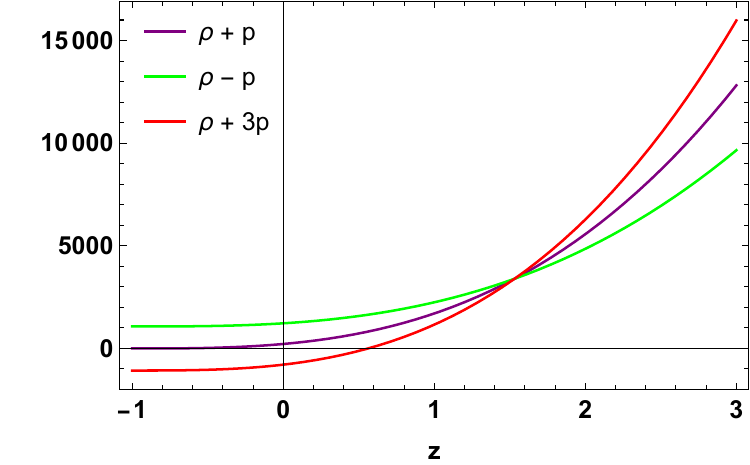}
	\caption{Above figure shows the behavior of $q(z)$ with the  constraint values of the cosmological free parameters}\label{ec}
\end{figure}

The energy conditions are fundamental constraints on the behavior of matter and energy in the universe, and are widely used in cosmological and theoretical physics applications. The mathematical formulations for the Null, Dominant, and Strong energy conditions are respectively derived as:
	\begin{small}
	\begin{equation}
		\rho + p  =C \left(H_0 h_1(z)\right)^{2-\Delta}-\frac{3 \alpha  H_0^2 \left(2 (\Omega_{0m}-1)+\beta \Omega_{0m}h_2(z) \right)}{4 \beta ^2+6 \beta +2} 
	\end{equation}
	
		\begin{equation}
		\rho - p  =C \left(H_0 h_1(z)\right)^{2-\Delta}+\frac{3 \alpha  H_0^2 \left(2 (\Omega_{0m}-1)+\beta \Omega_{0m}h_2(z) \right)}{4 \beta ^2+6 \beta +2}
	\end{equation}
	
		\begin{equation}
			\rho + 3 p  =C \left(H_0 h_1(z)\right)^{2-\Delta}-\frac{9 \alpha  H_0^2 \left(2 (\Omega_{0m}-1)+\beta \Omega_{0m}h_2(z) \right)}{4 \beta ^2+6 \beta +2}
	\end{equation}
\end{small}
where $h_1^2(z)=\Omega_{0m} (z (z+3)+3)z+1$ and $h_2=-2 + (3 + 3 z + 3 z^2 + z^3)$.\\
The energy conditions are shown in Fig. \ref{ec}, using the best values for the model's parameters. We found that:\\
- The Null Energy Condition (NEC) decreases over time in the early universe, but stays positive. However, it eventually becomes zero at late times.\\
- The Dominant Energy Condition (DEC) remains positive throughout the entire history of the universe, with no exceptions.\\
- The Strong Energy Condition (SEC) is followed in the early universe, but is violated at late times.

These results show that the energy conditions change over time and are important for understanding the universe's evolution.
\section{Conclusion}
In this study, we investigate the Barrow holographic dark energy model within the modified gravity framework of $f(Q,T)$ gravity, utilizing the Friedmann-Robertson-Walker (FRW) model to describe the universe's evolution. By exploring this new dark energy model, we aim to gain insights into the universe's expansion history and its dynamical evolution over time. Our analysis will provide a deeper understanding of the cosmos, The key features of the investigated model are as:

Our study reveals a significant phase transition in the deceleration parameter ($q$) from deceleration to acceleration at a redshift value of $z = 0.7413^{+0.0012}_{-0.0012}$. The present-day value of $q_0 = -0.601^{+0.0131}_{+0.0131}$ is in excellent agreement with recent cosmological observations, providing robust evidence for the accelerating expansion of the Universe. The transition redshift ($z_t$) and evolutionary behavior of $q$ are consistent with observations, marking a critical epoch in the Universe's evolution. Our findings validate the Barrow holographic dark energy model and reinforce the validity of the accelerating expansion of the Universe, consistent with a wealth of observational evidence.

The statefinder parameters $r$ and $s$ provide a valuable tool for distinguishing between various dark energy models, allowing for a quantitative and qualitative comparison of their evolutionary trajectories. The values of $r=1$ and $s=0$ serve as a benchmark, corresponding to the standard $\Lambda$CDM model, which accurately describes the universe's large-scale evolution. Any deviations from these values indicate a departure from the $\Lambda$CDM model, suggesting alternative dark energy scenarios. Our analysis of the statefinder parameters offers a deeper understanding of the cosmological model's dynamics, providing insights into the nature of dark energy and its role in shaping the universe's accelerating expansion. 

The $O_m(z)$ diagnostic parameter exhibits a constant value, indicating that the cosmological model under consideration is characterized by a flatness in the $O_m(z)$ curve. This constancy suggests that the model is consistent with the $\Lambda$CDM paradigm, where the dark energy density is proportional to the critical density. A constant $O_m(z)$ value also implies that the universe's expansion history is well-described by a single component, such as a cosmological constant, with no evidence for deviations or evolution in the dark energy's properties.\\
In conclusion, our comprehensive analysis of the $f(Q,T)$ gravity model has unveiled a rich and intricate portrait of the universe's evolution, seamlessly integrating theoretical predictions with cosmological observations. The model's energy density, pressure, and equation of state parameter trajectories collectively elucidate the universe's transition from a primordial singularity to an expanding cosmos, driven by dark energy's negative pressure. The evolutionary sequence of the equation of state parameter signifies a paradigm shift from deceleration to acceleration, with the contemporary value unequivocally indicating an accelerated expansion. Furthermore, the energy conditions' temporal evolution provides valuable insights into the universe's dynamics, underscoring the significance of these conditions in understanding the cosmos' evolution. Our findings collectively establish the $f(Q,T)$ gravity model as a robust and versatile theoretical framework, capable of accurately describing the universe's evolution across various epochs. This research contributes significantly to our understanding of the cosmos, offering a profound and detailed perspective on the universe's fundamental laws and mysteries.

\section*{Declaration of competing interest}
The authors declare that they have no known competing financial interests or personal relationships that could have appeared to influence the work reported in this paper.

\section*{Data availability}
No data was used for the research described in the article.

\section*{Acknowledgments}
The IUCAA, Pune, India, is acknowledged by the authors (S. H. Shekh \& A. Pradhan) for giving the facility through the Visiting Associateship programmes. Also, this research is funded by the Science Committee of the Ministry of Science and Higher Education of the Republic of Kazakhstan (Grant No. AP23483654).

\end{document}